\RequirePackage{fix-cm}
\documentclass[smallextended]{svjour3}       
\smartqed  

\usepackage{graphicx}
\usepackage{mathptmx}      
\usepackage[latin1]{inputenc}
\usepackage{amsmath}
\usepackage{amsfonts}
\usepackage{amssymb}
\usepackage{graphicx}
\usepackage{esint}
\usepackage{tikz}
\usepackage{tikz-3dplot}
\usepackage{array, caption, tabularx,  ragged2e,  booktabs}
\usepackage{subfigure}
\usepackage[10pt]{moresize}
\usepackage{hyperref}

\usetikzlibrary{shapes.geometric,calc}

\newcommand{\vect}[1]{{\mbox{\boldmath ${\mathit{#1}}$}}}

\renewcommand{\div}{\nabla\cdot}
\newcommand{\grad}{\nabla}
\newcommand{\curl}{\nabla\times}
\newcommand{\C}{{\bf C}}
\newcommand{\N}{{\bf N}}
\newcommand{\ie}{{\em i.e.} \/}
\newcommand{\eg}{{\em e.g.} \/}

 \journalname{The Journal of Engineering Mathematics}
\begin{document}

\title{The unstable temporal development of axi-symmetric jets of incompressible fluid}


\author{Earl S. Lester        \and
        Lawrence K. Forbes 
}


\institute{E. S. Lester \at
              Department of Mathematics and Physics \\
              University of Tasmania, Box 37, Hobart 7001, Australia \\
              \email{earl.sullivanlester@utas.edu.au}         
           \and
           L. K. Forbes \at
              Department of Mathematics and Physics \\
              University of Tasmania, Box 37, Hobart 7001, Australia \\
              Tel.: +61-3-6226 2720\\
              \email{larry.forbes@utas.edu.au}           
}

\date{Received: date / Accepted: date}

\maketitle

\begin{abstract}

We study the shear-driven instability, of Kelvin-Helmholtz type, that forms at the interface between a cylindrical jet of flowing fluid and its surroundings. The results of an infinitesimal-amplitude theory based on linearising the system about the undisturbed jet are given. A novel numerical model based on the Galerkin spectral method is developed and employed to simulate the non-linear development of the jet in the weakly-compressible Boussinesq regime. Representative results demonstrating the viability of this model are given, and are shown to agree with the linear analysis for small disturbances, but to develop cylindrical roll-up in the severely non-linear regime.


\end{abstract}

\section[]{Introduction}
\label{sec:intro}

In 1878, experimental observations and theoretical considerations led Lord Rayleigh to suggest two distinct mechanisms driving the instability and break-up of fluid jets \cite{rayleigh1878}. The first mechanism identified by Rayleigh is the well-known capillary instability, which is driven by surface tension  \cite{eggers2008} and readily observed on liquid-gas interfaces. The formation of capillary waves on liquid jets has been extensively studied, \eg \cite{osborneforbes2001}. The effects of gravity and surface tension on a liquid jet, the Rayleigh-Plateau instability \cite{eggers2008}, cause segmentation and drop formation. The second mechanism is ``of a more dynamical character" and ``operative even when the jet and its environment are of the same material" \cite{rayleigh1878}. Rayleigh showed that this is driven by the velocity shear between the jet and the environment, \cite{rayleigh1878}, the profile of this velocity shear determining the conditions for instability, \cite{drazinreid1981}.

The present article investigates this second mechanism with a focus upon the numerical simulation of the temporal development of the unstable jet and the resulting vortices. In this section some of the analyses and models of fluid jets and shear layers in past literature, and some of the relevant results of these, are briefly discussed, before outlining the structure and purpose of this article.

The stability characteristics of jet-like flows are investigated in the work of Batchelor and Gill, \cite{batchelorgill1962}, and the axi-symmetric equivalent of some well-known concepts in the stability of planar flow, such as Squire's theorem and Howard's semi-circle theorem, are developed \cite{batchelorgill1962}. The authors of \cite{batchelorgill1962} argue that the velocity of a jet leaving a circular orifice would take a ``top-hat" distribution, that is, uniform velocity with an abrupt velocity transition across the shear layer. As the jets progress spatially, the velocity profile flattens as the shear layer is thickened by the effects of viscous diffusion into a shallower ``far-downstream" profile \cite{batchelorgill1962}. Therefore, it is suggested that the spatial progression of a jet may be modelled by the choice of initial conditions to a temporally-developing jet. Support for this argument is found by Mattingly and Chang \cite{mattinglychang1974}, in a combined theoretical and experimental study. Anemometry is used to support the jet velocity profile evolution argued in \cite{batchelorgill1962}, and their theoretical results are confirmed. However, \cite{mattinglychang1974} argue the correct stability of this flow should be spatial and, furthermore, that a simple Galilean transformation between the two methods cannot be made. This is because, as Keller \textit{et al.} \cite{keller1973} note, Rayleigh's analysis implies that the oscillations of an unstable jet grow exponentially over its entire surface - this cannot be the case for a realistic jet, which oscillates at the nozzle only negligibly. 
	 
Following \cite{batchelorgill1962}, incompressible and unforced jets are simulated using spectral methods by Danaila \textit{et al.} \cite{danaila1997}, such that the initial velocity profile simulates the jets' spatial development. As was observed experimentally by \cite{becker1968}, the simulated jet of \cite{danaila1997} produces pairs of axially counter-rotating vortices. It is found that the most unstable mode of jet switches from sinuous (\ie helical) to varicose (\ie axial) as the Reynolds number of the flow increases beyond some critical value. This supports the notion that sinuous modes become dominant over the varicose modes as the shear layer thickens due to viscous diffusion as determined by \cite{batchelorgill1962}.
	 
Rising columns of fluid surrounded by a denser ambient fluid, \ie buoyant plumes, are closely related to fluid jets. These have been theoretically and numerically investigated by Hocking and Forbes \cite{hockingforbes2010} for steady flows and Forbes and Hocking \cite{forbeshocking2013} and Letchford \textit{et al.} \cite{letchfordforbes2012} in unsteady flows. These works develop numerical spectral methods for viscous flows approximated using Boussinesq theory. This latter work accounts for turbulent flows by an entrainment approximation. Viscous entrainment is found to cause a widening of the plume as it rises due to the generation of vortices, while without entrainment the jet grows increasingly narrow \cite{letchfordforbes2012}. The initial conditions of the simulations were chosen so as to allow the viscous jet to mimic an inviscid jet model, facilitating some comparison between the two, at least for early times preceding the formation of a curvature singularity on the inviscid interface \cite{forbeshocking2013}.

If the K-H instability is assumed to be the fundamental mechanism behind observed jet phenomena then the well-studied, classical planar shear layer is relevant. Vortex sheets are the asymptotic limit of an infinitesimally thin shear layer, \cite{drazinreid1981}. If surface tension and gravity are neglected, classical theory predicts the vortex sheet is unstable \cite{chandra1981}. Although certain stable solutions may be found for some ranges of wave-numbers of the disturbance when gravity or surface-tension are included, Siegel \cite{siegel1995}, suggests that surface tension cannot completely suppress the formation of a singularity in the curvature of the vortex sheet which will occur at a finite, critical time, as shown by Moore \cite{moore1979}. Furthermore, the growth rate is found to increase linearly with respect to the wave-number, so higher modes of vibration are more unstable, \cite{batchelor1970}. This suggests the ideal model is innately ill-posed, \cite{krasny1986}.

Chorin and Bernard in \cite{chorin1973} used a ``vortex blob" method to simulate solutions, which Krasny \cite{krasny1986} showed continues to yield a solution after Moore's critical time. The curves generated by this method are shown to approach the same classical solution of the idealised vortex sheet but with spiral details dependent on the blob size. Alternatively, Baker in \cite{baker1990} replaced the vortex sheet by a layer of finite thickness and uniform vorticity, arguing that this models thin, viscous shear layers. This is suggestive of the prior work of Drazin \cite{drazin1958} in which the stability of a flow with a smooth and continuous transition in velocity and density is studied. The profiles are assumed to have the form of a hyperbolic tangent function. The effect of this is found to be stabilising for a range of wave-numbers, \cite{drazin1958}. 

Tryggvason \textit{et al.} \cite{tryggvasondahmsbeih1991} consider viscosity as a regularising agent. The viscous model deviates from the idealised model, particularly in the detail of the core structure of the spirals, when the sheet has a finite initial thickness; although it is noted that the simpler dynamics of the inviscid model should only be expected to simulate the large scale features, \cite{tryggvasondahmsbeih1991}. The problem is similarly approached by Chen and Forbes \cite{chenforbes2011}, in which the inviscid vortex sheet was studied by spectral numerical methods and compared to a viscous and weakly-compressible version, simulated by finite difference methods. The inviscid model showed points of curvature extrema approach one another as the curvature singularity forms, before any obvious sign of interface roll-over. An unstable viscous shear layer is modelled and shown to reproduce the cat's-eye spirals effectively. 
	 
The focus of the present article is the temporal instability of a jet of fluid interpreted as a cylindrical vortex sheet. Section \ref{sec:idealised_model} introduces the first classical model which uses linearisation to investigate the evolution of a discontinuous interface of an idealised jet. Section \ref{sec:boussinesq_model} develops a viscous model based upon the weakly-compressible Boussinesq approximation. The discontinuous interface assumption is relaxed and an inhomogeneous, one-fluid model with a inter-facial zone is employed. Due to the non-linear complexities of this system, a spectral scheme is developed. The numerical scheme is novel in that it a takes a simple and intuitive approach to approximating the non-linear model following global Galerkin computational methods. The simulation was able to reproduce the expected structure of the K-H instability with high accuracy and low computational costs. Results demonstrating the viability of this numerical model are given in Section \ref{sec:results}, in which a comparison is made between the two models in the appropriate limits of parameters showing the correspondence of the novel model with expected behaviour; the effect of initial conditions, discontinuous and smoothed hyperbolic tangent profiles are investigated. These results are illustrated by considering time series of contours of density and scalar vorticity. The expected axi-symmetric over-turning waves are formed along the inter-facial region. Following a discussion of these results, the paper concludes in Section \ref{sec:conclusion}.

\section[]{The Inviscid Model}
\label{sec:idealised_model}

This section presents a preliminary investigation of the cylindrical Kelvin-Helmholtz instability. The model studied aims at capturing the key aspects of the shear driven instability of a cylindrical free-surface such that classical results may be recovered.

Let a column or cylinder of fluid be characterised by velocity $\vect{q}_{\tiny{1}}$ and density $\rho_{1}$, such that axial symmetry and periodicity holds, the axis aligned with the axial component of a system of cylindrical-polar coordinates ($r,\theta,z$). The cylinder is surround by another external fluid of perhaps different $\vect{q}_{\tiny{2}}$ and $\rho_{2}$. If fluid speeds are not comparable to the speed of sound of the fluid, \cite{batchelor1970} we may assume the effects of compressibility are negligible so that the densities $\rho_{1}$ and $\rho_{2}$ are constant in each fluid. If the properties differ then it is possible to distinguish an interface $r = \eta(z,t)$ between the fluids which are immiscible but lack surface tension. 

Inviscid theory makes a good approximation to real fluid flow if shear is stronger than viscous forces (\ie the Reynolds number $Re \gg 1$), \cite{batchelor1970}, when accompanied by the appropriate initial and boundary conditions. Regions where the theory fails are often boundary layers, since viscous fluids must satisfy a no-slip condition on rigid boundaries, \cite{batchelor1970}. The present model has no rigid boundaries and therefore suggests this idealisation, which is further validated by observations of the Kelvin-Helmholtz instability occurring in inviscid super-fluids, \cite{andersson2003}. 

Body forces on the fluids would likely be gravity directed down along the $z$-axis as in the case of a rising/falling stream or directed radially inwards as could be the case of a self-gravitating astrophysical object. For the present analysis, however, we assume them to be negligible -- perhaps corresponding to a laboratory experiment performed on a small length-scale or fast time-scale, as is often the case in fluid dynamical experiments, \cite{forbes2011cylindrical}. 

The initial conditions of this model are taken to be irrotational such that, by the Kelvin's circulation theorem, the vorticity $\vect{\zeta} = \vect{0}$ throughout each region, for all time. Note that $\vect{\zeta} = \omega\,\hat{\vect{e_{\theta}}}$  due to the axi-symmetry. Irrotationality is identically satisfied by a scalar velocity potential defined by $\vect{q} = \grad\Phi$. The continuity of an incompressible fluid requires $\div \vect{q} = 0$, thus the velocity potential must satisfy the Laplace equation $\nabla^{2} {\Phi} = 0$.

Suppose a steady-state of this system is given by $\vect{q}_{i} = c_{i} \hat{\vect{e_{z}}}$ for $ i = 1,2$, representing the internal and external fluids respectively. Suppose, too, that the interface has mean radius $R$, so that $\eta = R$. Here, $c_{1},c_{2}$ and $R$ are constants and $R > 0$, which corresponds to the case where the internal fluid column is a perfect cylinder of constant radius $R$, flowing with constant speed $c_{1}$ in the axial or $z$ direction, and the external fluid has constant speed $c_{2}$.

Finally, adopting dimensionless variables, the  wave-number $\lambda / {2 \pi}$ of the interface disturbance is taken as the length scale while $c_{1}$ and $\rho_{1}$ of the internal fluid are chosen to scale speed and density respectively.

\subsection[]{Boundary Conditions}
\label{subsubsec:bcs_1}

Certain boundary conditions must be satisfied on the interface $r = \eta(z,t)$, along the cylinder axis, $r = 0$, and in the far-field away from the jet, \ie as $r \rightarrow \infty$. 

The first inter-facial boundary condition is the kinematic condition, \cite{drazinreid1981}, which can be expressed as	
\begin{equation}
\label{eq:kinematic_BC}
u_{i} = \dfrac{\partial \eta}{\partial t} + v_{i}\dfrac{\partial \eta}{\partial z} 	\quad \quad , \quad i = 1,2,
\end{equation}
on $r = \eta(z,t)$, where $u_{i}$ and $v_{i}$ are the radial and axial components of the velocity $\vect{q}_{i}$, respectively.

Secondly, a dynamic boundary condition must be enforced which requires the continuity of normal stresses of fluid along the interface \cite{drazinreid1981}. For the idealised Eulerian system currently considered, energy conservation, expressed as the Bernoulli equation, allows the pressures to be eliminated and the dynamic condition becomes, on $r = \eta(z,t)$
\begin{equation}
\label{eq:dynamic_BC}
\delta \Big( \dfrac{\partial \Phi_{2}}{\partial t} + \dfrac{1}{2}(u_{2}^2 + v_{2}^2) \Big) - \Big( \dfrac{\partial \Phi_{1}}{\partial t} + \dfrac{1}{2}(u_{1}^2 + v_{1}^2) \Big)	= \dfrac{1}{2} \Big( \delta \gamma^2 - 1 \Big) 
\end{equation}
where $\gamma = c_{2}/c_{1}$ and $\delta = \rho_{2}/\rho_{1}$. The right-hand-side of (\ref{eq:dynamic_BC}) is determined by the requirement that these boundary conditions satisfy the steady-state described above.

Further conditions on the boundaries of the flow domain must be specified.  On the $z$-axis,
\begin{equation}
\label{bc:on_axis}
u_{1} = 0	\quad , \quad	\dfrac{\partial v_{1}}{\partial r} = 0 \quad, \quad r = 0 \text{.} 
\end{equation}

In the outer fluid, at large radial distances
\begin{equation}
\label{bc:far_field}
u_{2} \rightarrow 0	\quad , \quad	v_{2} \rightarrow c_{2} \quad , \quad \dfrac{\partial v_{2}}{\partial r} \rightarrow 0 \quad, \quad \text{as} \quad r \rightarrow \infty \text{.} 
\end{equation}

Condition (\ref{bc:on_axis}), required by the assumed axial symmetry of the flow, avoids flow singularities on the axis for the internal fluid $i = 1$, and (\ref{bc:far_field}) ensures flow uniformity of the outer fluid in the far-field.

Finally, the flow disturbance is assumed to be axially periodic with a (dimensionless) period of $2 \pi$; therefore, any  solution function and all its derivatives must match at the boundaries of each $2 \pi$ period.

\subsection[]{Analysis}
\label{sec:idealised_model_analysis}

Let the functions describing the interface and velocity potential be approximated as perturbations of order $\epsilon$ around the steady-state suggested above, now non-dimensional such that the undisturbed jet has radius $a = (2\,\pi\,R)/\lambda$ and flows axially and axi-symmetrically in an external medium of speed $\gamma$. We seek the first-order unknown $P_{i}(r,z,t)$ and $ H(z,t)$ perturbations to the velocity potentials and interface respectively. The validity of this approximation requires the amplitude of the initial disturbance, $\epsilon$, to be infinitesimally small (in relation to the other scale factors).

The incompressible continuity equation and the linearity of the Laplacian, gives Laplace equations on each $P_{i}$. Separation of variables, in the cylindrical and axially $2\pi$-periodic region with boundary conditions (\ref{bc:on_axis}) and (\ref{bc:far_field})and physical conditions, suggests the solution
\begin{equation}
\big( P_{1}(r,z,t), P_{2}(r,z,t),H(z,t) \big) =  \big( A\,I_{0}(n\,r),B\,K_{0}(n\,r),C \big)\,\textrm{e}^{s\,t + \textrm{i}\,n\,z} \nonumber
\end{equation}
where $\{A, B, C\} \in \C$ are constants, $I_{0}$ and $K_{0}$ are the zeroth-order modified Bessel functions of the first and second kind, \cite{abramowitz1964}, $n \in \N$ the wave-number and $s \in \C$ is the growth rate. Thus, if $Re \{ s \} > 0$ then the waves are unstable.

The unknown functions evaluated on the interface $r = \eta(z,t)$ are projected onto the known, steady interface $r = a$. Under this approximation, substituting in the first-order perturbation expansion, the interface boundary conditions, (\ref{eq:kinematic_BC}) and (\ref{eq:dynamic_BC}) become, on $r = a$,
\begin{align}
\label{eq:Linear_interface_BC}
\dfrac{\partial P_{1}}{\partial r} &= \dfrac{\partial H}{\partial t} + \dfrac{\partial H}{\partial z}  \nonumber \\
\dfrac{\partial P_{2}}{\partial r} &= \dfrac{\partial H}{\partial t}  + \gamma \dfrac{\partial H}{\partial z}	\\
\delta\,\big( \dfrac{\partial P_{2}}{\partial t} &+ \gamma \dfrac{\partial P_{2}}{\partial z} \big) = \dfrac{\partial P_{1}}{\partial t}  + \dfrac{\partial P_{1}}{\partial z}	\text{.} 	\nonumber
\end{align}

On evaluation of the system of boundary conditions for the assumed form of the functions, and elimination of constants, the unknown $s$ is determined to be
\begin{equation}
s = \pm \sigma - \textrm{i}\,\omega		\nonumber
\end{equation}
with growth rate
\begin{equation}
\label{eq:growth_rate}
\sigma = \dfrac{n\,(\gamma - 1)\,\sqrt{\delta\,\hat{I}_{n}\, \hat{K}_{n}}}{[\hat{I}_{n} + \delta\,\hat{K}_{n}]} 
\end{equation}
and angular frequency	
\begin{equation}
\label{eq:angular_freq}
\omega = \dfrac{n\,[\hat{I}_{n} + \delta\,\gamma\,\hat{K}_{n}] }{[\hat{I}_{n} + \delta\,\hat{K}_{n}]} \text{,}
\end{equation}
where 
\begin{equation}
\hat{I}_{n} = \dfrac{I_{0}(n\,a)}{I_{1}(n\,a)} \quad	\text{and} \quad \hat{K}_{n} = \dfrac{K_{0}(n\,a)}{K_{1}(n\,a)} \text{.}
\end{equation}

The growth rate, equation (\ref{eq:growth_rate}), immediately suggests several results. Firstly, any velocity shear, such that $\gamma \neq 1$, causes the interface to be unstable, and the larger the magnitude of the shear, the greater the instability. Secondly, a decrease in the radius of the jet, $a$, produces a more unstable jet, and similarly, disturbances of higher wave-numbers, $n$, are more unstable. This is confirmed by considering the limit $(n\,a) \gg 1$. By making use of the asymptotic behaviour of the modified Bessel function for large argument (as given in \cite{abramowitz1964}), equations (\ref{eq:growth_rate}) and (\ref{eq:angular_freq}) simplify, suggesting that disturbance travels with the mean of the two fluid speeds and grows at a rate of half the fluid speed differences. This corresponds to physical intuition as well as the results of the classical planar theory, \cite{drazinreid1981}.

\section[]{The Boussinesq Model}
\label{sec:boussinesq_model}

In the second model, the two-fluid system separated by an idealised interface of the inviscid model in Section \ref{sec:idealised_model} is replicated by a single-fluid system with density and velocity varying over an inter-facial zone of finite width. This single-fluid system can be viably approximated by Boussinesq theory when the density variation is small. This approximation is made in Section \ref{sec:boussinesq_governing equations}, such that the model becomes one of a viscous and weakly-compressible fluid. The assumptions of axial periodicity and axi-symmetry are retained. A system of governing equations and boundary conditions is derived, to which a numerical approximation is proposed. The numerical scheme adopted is a Galerkin spectral method and is developed in Section \ref{sec:numerical_scheme}. 

\subsection[]{The Governing Equations}
\label{sec:boussinesq_governing equations}

The Boussinesq theory supposes, firstly that variations of density are continuous and that the effects of pressure on density are negligible, \cite{drazinreid1981}. Secondly, this variation may be ignored except when the density variation is coupled with an external force, as in buoyancy effects, which may be of significant order, \cite{spiegel1960}. 

The density is assumed to be approximated by a constant base density that is slightly perturbed by a variable density which is much smaller than the base field. Assuming the key dimensional quantities to be scaled by the same factors as for the previous model, the dimensionless density field is taken to be 
\begin{equation}
\label{eq:bq_density_approx_nd}
\rho(r,z,t) = 1 + \overline{\rho}(r,z,t)
\end{equation}
such that $\overline{\rho} \ll 1$, which suggests that, on substitution of (\ref{eq:bq_density_approx_nd}), the continuity equation would operate on two terms of significantly different order and as such may be treated as two independent equations. Due to the constancy of the background density field, the incompressible continuity equation still holds but is now supplemented by a convection-diffusion equation governing the density perturbation, namely
\begin{equation}
\label{eq:governing_density_nd}
\dfrac{\partial \overline{\rho}}{\partial t}  + u \dfrac{\partial \overline{\rho}}{\partial r} + v \dfrac{\partial \overline{\rho}}{\partial z} =  \theta \,\nabla^{2} \overline{\rho}
\end{equation}
where $\theta$ is a dimensionless measure of the density perturbation diffusion.

The fluid is assumed to be Newtonian but not inviscid such that the flow is governed by Cauchy's momentum equation. Upon application of the Boussinesq approximation, the governing momentum equation may be recast (by curl operation) in terms of the (scalar) vorticity--stream-function ($\omega$-$\psi$) formulation which identically satisfies the continuity equation and eliminates pressure terms therefore simplifying the computation, \cite{hussainizang1987}. The governing momentum equation becomes 
\begin{equation}
\label{eq:governing_vorticity_nd}
\dfrac{\partial \omega}{\partial t} + u \dfrac{\partial \omega}{\partial r} + v \dfrac{\partial \omega}{\partial z} - \dfrac{u \, \omega}{r}  = \dfrac{1}{Re}\, \Delta \omega +  \dfrac{1}{Fr}\,\dfrac{\partial \overline{\rho}}{\partial r}	\text{.}
\end{equation}
Here, $\omega$ is the scalar vorticity, $\Delta \equiv  (\nabla^{2} \, - 1/r^2 )$ and two dimensionless numbers have been introduced: a Reynolds number, $Re = (\lambda c_{1} \rho_{1})/(2 \pi \mu)$, where $\mu$ is the viscosity, and a Froude number, $Fr = (2 \pi c_{1}^{2})/(g \lambda)$ in which we have assumed a downwards (negative $z$) gravitational acceleration of constant $g$. Also note that the scalar vorticity satisfies $\omega = - \Delta \psi$ where $\psi$ is a stream-function defined by $\vect{q} = \curl{(\psi\,\hat{\vect{e_{\theta}}})}$.

Equation (\ref{eq:governing_vorticity_nd}) is the vorticity equation, \cite{batchelor1970}, and in conjunction with equation (\ref{eq:governing_density_nd}) and the above velocity-vorticity-stream-function relations, with appropriate boundary and initial conditions, determines the flow.

\subsection[]{The Boundary Conditions}
\label{subsubsec:bcs_2}

In this model the interface is replaced by an inter-facial zone, the structure of which will be imposed by the initial conditions. Thus there are no boundary conditions to apply on a moving inter-facial surface of density discontinuity. However, the system proposed by this model will be solved numerically and therefore boundary conditions arise as artifices of the numerical simulation. That is, an artificial boundary must be imposed to replicate the far-field conditions at infinity. The conditions at this boundary must have a minimal impact on the model. The condition here is taken to be a slip condition, that is, such that the fluid may flow unhindered parallel to some computational boundary at $r = R_{\infty}$, with no material transport across this. Therefore, boundary conditions must be imposed on the $z$-axis and on the artificial computational boundary. In order to mimic the conditions (\ref{bc:on_axis}) and (\ref{bc:far_field}) of the inviscid model of Section \ref{sec:idealised_model}, the boundary condition are thus
\begin{equation}
\label{bc:boussinesq_bcs}
\omega = 0 \quad \text{,} \quad \dfrac{\partial \overline{\rho}}{\partial r} = 0 \quad \text{on} \quad r = 0 \quad \text{and} \quad r = R_{\infty} \,\text{.}
\end{equation}

\subsection[]{The Spectral Method}
\label{sec:numerical_scheme}

Computational Galerkin methods seek an approximate solution to the system as an expansion in basis functions that are global, independently satisfying all homogeneous boundary conditions \cite{canuto2012}. By an appropriate choice of basis functions, the approximation converges as the number of modes in the series becomes infinite. In implementation the series must terminated at a finite mode; thus the accuracy of the approximation is determined by the rate of convergence of the series. Completeness of the set of basis functions and smoothness of the target solution ensures the exponential convergence of the approximation \cite{boyd2001}. 

Interpolation errors, whereby high wave-numbers are interpreted as effectively lower due to insufficient sampling, were found to be avoided by taking the number of nodes (or points) in one dimension, $n$, as $n \geq 5\,N$, where $N$ represents the number of modes of the series representation in that dimension. The two spatial dimensions $r$ and $z$ were generally taken to have the same number of nodes and modes. The highest mode or truncation term of the series, represented by $N$, determines the accuracy of the simulation. However, if there is strong convergence to the true solution then beyond some threshold truncation, higher modes are redundant. Due to the initially discontinuous axial velocity and the Gibbs phenomenon this causes \cite{kreyszig2010}, the convergence rate is not expected to be exponential, but a truncation of $N = 41$ was sufficient, within graphical precision, producing results within a computationally manageable time. 

Solving Helmholtz equations with boundary conditions (\ref{bc:boussinesq_bcs}) by separation of variables produces Sturm-Liouville problems which suggest that an appropriate series approximation for the present problem is
\begin{equation}
\label{eq:psi_rep_approx}
\psi_{\scriptscriptstyle{MN}}(r,z,t) = \psi_{\scriptscriptstyle{\textit{b.g.}}} + \sum_{m = 1}^{M} \sum_{n = 0}^{N} J_{1}(\alpha_{m} r) \big[A_{mn}(t)\cos(n\,z)	+ B_{mn}(t)\sin(n\,z) \big]
\end{equation}
where $\psi_{\scriptscriptstyle{\textit{b.g.}}}$ is a known stream-function that satisfies boundary conditions for a background field. The quantity $J_{1}(x)$ represents the Bessel function of order 1, and $j_{1,m}$  $m = 1,2,...$ are its zeros, following \cite{abramowitz1964}. The remaining constants in the expression (\ref{eq:psi_rep_approx}) are $\alpha_{m} = (j_{1,m}/R_{\infty})$, and the time-dependent Fourier coefficient functions $A_{mn}(t)$ and $B_{mn}(t)$ are to be determined.

A similar series representation to (\ref{eq:psi_rep_approx}) is adopted for $\overline{\rho}$ with, however, the zero-th order Bessel function $J_{0}(x)$ replacing $J_{1}(x)$, and new coefficient functions $C_{mn}(t)$ and $D_{mn}(t)$. The approximate vorticity $\omega_{\scriptscriptstyle{MN}}$ and velocity components $u_{\scriptscriptstyle{MN}}$ and $v_{\scriptscriptstyle{MN}}$ are determined by direct differentiation of (\ref{eq:psi_rep_approx}).

Fourier analysis of the governing equations (\ref{eq:governing_density_nd}) and (\ref{eq:governing_vorticity_nd}), upon substitution of the truncated series, produces a set of ODEs governing the time-dependent coefficients, for $m = 1,...,M$ and $n = 0,1,..,N$. This system can be written
\begin{align}
\label{eqs:coeff_ode_system}
\dfrac{d A_{mn}}{d t} & = - \,\dfrac{k_{mn}^2}{Re} A_{mn} \, - \dfrac{c_{n}}{\nu_{mn}^2} \, I^{1}_{mn}	\nonumber \\
\dfrac{d B_{mn}}{d t} & = - \,\dfrac{k_{mn}^2}{Re} B_{mn} \, - \dfrac{c_{n}}{\nu_{mn}^2} \, I^{2}_{mn}	\nonumber	\\	
\dfrac{d C_{mn}}{d t} & = - k_{mn}^2 \, \theta \, C_{mn} \,  - \dfrac{c_{n}\,k_{mn}^2 }{\nu_{mn}^2} \, I^{3}_{mn}	\nonumber	\\
\dfrac{d D_{mn}}{d t} & = - \,  k_{mn}^2  \theta\, D_{mn} \, - \dfrac{c_{n}\,k_{mn}^2}{\nu_{mn}^2} \, I^{4}_{mn}		
\end{align}	
with constants defined as
\begin{equation}
k_{mn}^2 = \alpha_{m}^2 + n^2 \text{,}
\end{equation}
\begin{equation}
\nu_{mn} =  \sqrt{\pi} \, k_{mn} \, R_{\infty} \, J_{0}(j_{1,m})	\text{,}
\end{equation}
\begin{equation}
c_{n} = \begin{cases}
1 \,\, & \text{for} \quad n = 0	\\
2 \,\, & \text{for}  \quad n \neq 0 \text{.}
\end{cases} 
\end{equation}	

The integrals over the entire spatial domain are	
\begin{align}
\label{eqs:integrals}
I^{1}_{mn} = \int_{0}^{R_{\infty}} r J_{1}(\alpha_{m} r) \int_{-\pi}^{\pi} H_{\scriptscriptstyle{MN}} \cos(n z) \,dz\,dr	\nonumber \\
I^{2}_{mn} = \int_{0}^{R_{\infty}} r J_{1}(\alpha_{m} r) \int_{-\pi}^{\pi} H_{\scriptscriptstyle{MN}} \sin(n z) \,dz \,dr \nonumber \\
I^{3}_{mn} = \int_{0}^{R_{\infty}} r J_{0}(\alpha_{m} r) \int_{-\pi}^{\pi} J_{\scriptscriptstyle{MN}} \cos{(n z)}  \,dz\,dr \nonumber \\
I^{4}_{mn} = \int_{0}^{R_{\infty}} r J_{0}(\alpha_{m} r) \int_{-\pi}^{\pi} J_{\scriptscriptstyle{MN}} \sin{(n z)}  \,dz\,dr	\text{.}
\end{align}	

In these expressions, the non-linear complexities of these equations are nested in the terms $H_{\scriptscriptstyle{MN}}$ and $J_{\scriptscriptstyle{MN}}$, given by
\begin{equation}
\label{eq:hmn}
H_{\scriptscriptstyle{MN}} \equiv u_{\scriptscriptstyle{MN}} \dfrac{\partial \omega_{\scriptscriptstyle{MN}}}{\partial r} +
v_{\scriptscriptstyle{MN}} \dfrac{\partial \omega_{\scriptscriptstyle{MN}}}{\partial z} - \dfrac{u_{\scriptscriptstyle{MN}} \, \omega_{\scriptscriptstyle{MN}}}{r}  - \dfrac{1}{Fr}\,\dfrac{\partial \overline{\rho}_{\scriptscriptstyle{MN}}}{\partial r}
\end{equation}
and
\begin{equation}
\label{eq:jmn}
J_{\scriptscriptstyle{MN}} \equiv u_{\scriptscriptstyle{MN}} \dfrac{\partial \overline{\rho}_{\scriptscriptstyle{MN}} }{\partial r} + v_{\scriptscriptstyle{MN}} \dfrac{\partial \overline{\rho}_{\scriptscriptstyle{MN}} }{\partial z}	.
\end{equation}

The system, (\ref{eqs:coeff_ode_system}), of $2\,M\,(2\,N + 1)$ ODEs, is given an initial state, introduced and discussed in Section \ref{sec:results}, and time-marched forward by a Runge-Kutta-Fehlberg method, see \textit{e.g.} \cite{atkinson2008}. At each time-step, the approximated variables are reconstructed and $H_{\scriptscriptstyle{MN}}$ and $J_{\scriptscriptstyle{MN}}$ are formed by (\ref{eq:hmn}) and (\ref{eq:jmn}). The integrals (\ref{eqs:integrals}) are evaluated by a Legendre-Gauss quadrature routine, based upon a program developed by G. von Winckel \footnote{Available freely online from \url{https://au.mathworks.com/matlabcentral/fileexchange/4540-legendre-gauss-quadrature-weights-and-nodes?s_tid=prof_contriblnk}}. 

\subsection[]{The Initial Conditions}
\label{subsubsec:ics}

In order to facilitate a comparison between the two models presented, the initial conditions must be matched. Consider the idealised model presented in Section \ref{sec:idealised_model} and suppose the interface perturbation is $H(z,0) = \cos{(z)}$ with the first time derivative $\partial H/\partial t = 0 $ at the initial instant $t = 0$. These initial conditions and the relations (\ref{eq:Linear_interface_BC}) , suggest that an initial condition for the axial velocity component is
\begin{equation}
\label{eq:discontinuous_velocity_ic}
v(r,z,0) = \begin{cases}
1 -  \epsilon \dfrac{I_{0}(r)}{I_{1}(a)} \cos{(z)} \, \text{,} \quad \quad \quad \,\, 0 < r < a &	\\
\gamma (1 +  \epsilon \dfrac{K_{0}(r)}{K_{1}(a)} \cos{(z)}) \, \text{,} \quad \,\,\,\,\, a < r < R_{\infty} &
\end{cases}
\end{equation}
at $t = 0$. Note that the idealised model has a discontinuous density across the interface $r = a$ at this time. These discontinuities conflict with the assumptions of the Boussinesq theory, and produce Gibbs phenomenon in the Fourier series approximation and furthermore cause the convergence to be $O(1/N)$ where $N$ is the maximum mode, \textit{i.e.} no longer exponential, \cite{kreyszig2010}. Therefore, following \cite{drazin1958}, \cite{michalke1984} and \cite{batchelorgill1962}, among others, the second set of initial conditions approximate the interface as a shear layer and smooth stratification, an inter-facial zone. 

The adopted profiles take a hyperbolic tangent form, such that the second initial condition considered is
\begin{equation}
\label{eq:continuous_velocity_ic}
v(r,z,0) = 1 + \dfrac{\gamma - 1}{2} \big(1 + \tanh(S_{1}[r - \eta(z,0)]) \big)
\end{equation}
where the parameter $S_{1}$ determines the gradient of the initial transition and thickness of the inter-facial zone. A similiar profile is adopted for a smooth stratification of the density perturbation, the relevant parameter in that case being the density ratio $\delta$, with $S_{2}$ as the steepness parameter. Initial conditions can thus be supplied to the system of ODEs (\ref{eqs:coeff_ode_system}), by direct Fourier analysis of the starting speed (\ref{eq:discontinuous_velocity_ic}) or (\ref{eq:continuous_velocity_ic}) and the corresponding initial density profile.

\section[]{Results and Discussion}
\label{sec:results}

This section presents and discusses some results of simulations produced by the numerical scheme developed in the previous section, with the intent of validating the accuracy of the numerical scheme, using comparisons with the linearised and inviscid models. Once this correspondence is established, we investigate the differences of the discontinuous and smoothed axial velocity initial conditions given by Equations (\ref{eq:discontinuous_velocity_ic}) and (\ref{eq:continuous_velocity_ic}) respectively. 

The presentation of these results has the intended purpose of demonstration and affirmation of the developed model, and as such the study has been confined to a small part of the parameter space. A preliminary exploration of the effects of viscosity is given, but the results remain focused upon the validity of the model. The effect of a flow-directed gravity-like body force that is allowed for by this model is not investigated here.

\subsection[]{Comparison of Models}
\label{subsec:comparison}

The validity of the models developed in this article can be supported by comparing the results of the linear, analytic model and the non-linear, viscous model in the appropriate limit of the parameters to approximate an inviscid and incompressible flow. This comparison requires matching the initial conditions of the two models and the set of common model parameters (these being the speed and density ratios $\gamma$ and $\delta$, the initial disturbance amplitude $\epsilon$ and the initial mean jet radius $a$) as closely as possible. The effect of these parameters shared by the two models, on the stability and development of the flow were investigated but are not included in this presentation. 

The correspondence of the two models is only expected in the very early times of the simulation since the linear theory neglects the influence of finite-amplitude effects which are expected to become significant quickly. Once a correspondence between the models has been qualitatively established for these early times and range of parameters, the viscous, numerical model can be used to explore the non-linear and finite amplitude effects, which were beyond the domain of validity of the inviscid, analytic model. 

\begin{figure}
 \includegraphics[width=1\columnwidth, scale=1]{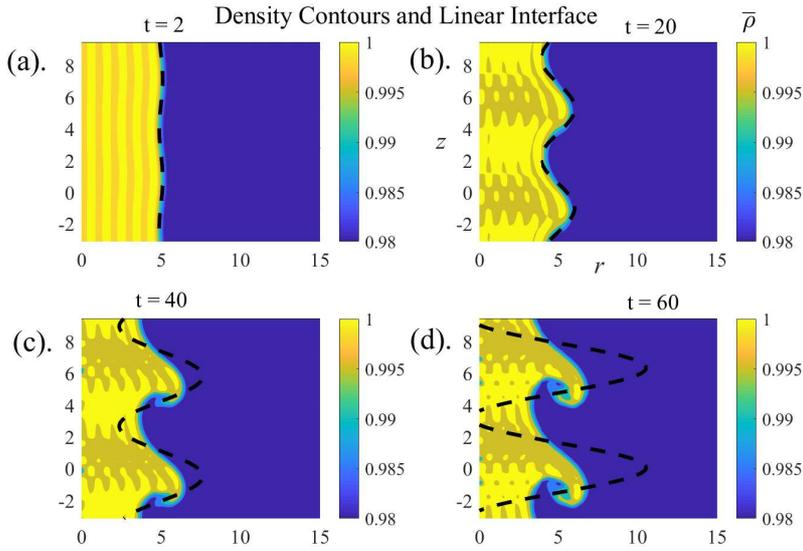}
\caption{Density contours presented at several times primarily to demonstrate the initial correspondence of the idealised (the exact interface solution of which is represented by the black dashed line) and Boussinesq models (of which colour-filled density contours are shown) and the increasing deviation in time of the two models. The parameters used for these simulations are: $N = 41$, $n = 255$, $\gamma = 0.93$, $\delta = 0.98$, $a = 5$, $\epsilon = 0.05$, $S_{2} = 8$, $1/Fr = 0$, $1/Re = 10^{-5}$, $\theta = 10^{-5}$.}
\label{fig:density_present_t_30_120}       
\end{figure}

In Figure \ref{fig:density_present_t_30_120}, this comparison is made. A sequence of four density profiles is presented for a typical case, here at four different dimensionless times $t = 2, 20, 40, 60$. These densities $\overline{\rho}$ were computed using the spectral approach described in Section \ref{sec:numerical_scheme}. Superposed on each picture, as a heavier dashed line, is the prediction for the interface profile from the linearised inviscid theory of Section \ref{sec:idealised_model_analysis}.

It is readily observed in Figure \ref{fig:density_present_t_30_120} that the linear interface and the central region of the inter-facial zone initially show very close agreement in the early times of the development of the instability. Despite the initiation of non-linear effects intimated by the slight asymmetry observed of the inter-facial wave at $t = 20$ there is still a strong correspondence between the interfaces of the two models, suggesting finite-amplitude effects are not yet significant. However, beyond this time, as shown by the times $t = 40$ and $t = 60$, the two models show significant deviation in results. The phases of the waves are in agreement at $t = 40$, but there is no longer a correspondence between the linear interface and the centre of the inter-facial zone (which may be taken to approximate an interface). At these times, the shear has begun to cause the expected non-linear rolling-over of the finite-amplitude protrusion of the perturbed interface unto itself, characteristic of the K-H instability, which the linear model is incapable of capturing. The plots suggest a slight thickening of the inter-facial zone in regions of high non-linearity. The time $t = 60$ shows the non-physical end-result of the linear model, where the unchecked exponential growth of the interface has caused it to cross the axis $r = 0$. Also near the axis, at this time, the density contours show a slight decrease, which we suspect is due to numerical noise (as partially intimated by Figure \ref{fig:2:vorticity_comparison}).
 
\subsection[]{Comparison of Initial Conditions}

\begin{figure*}
	\includegraphics[width=\columnwidth, scale=1]{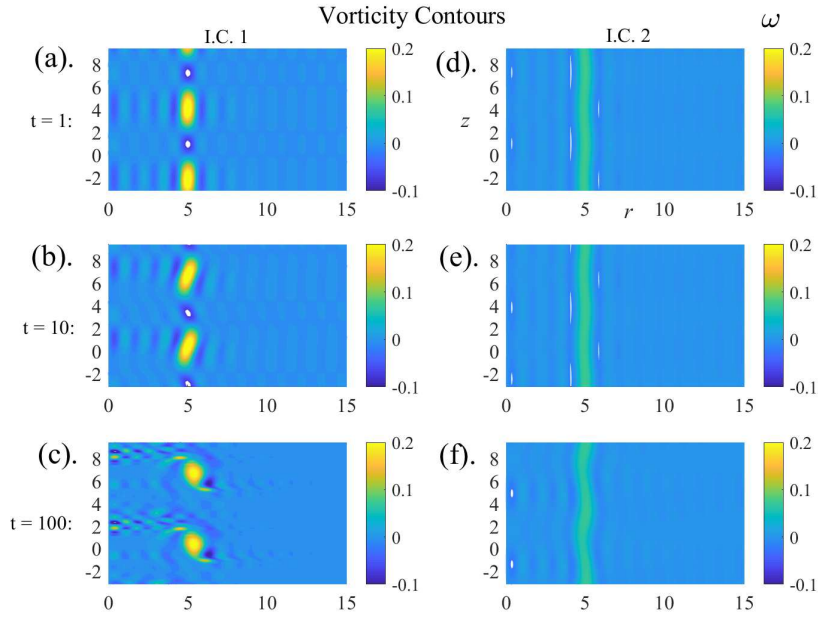}
	\caption{Vorticity contours taken at $t = 1$, $t = 10$ and $t = 100$ simulation units, showing the effect of the choice of initial condition, either discontinuous axial velocity (``I.C. 1" given by Equation (\ref{eq:discontinuous_velocity_ic}) as determined by the linear theory or continuous hyperbolic tangent axial velocity (``I.C. 2" given by Equation (\ref{eq:continuous_velocity_ic})). The parameters otherwise are identical. They are: $N = 31$, $n = 155$, $\gamma = 0.95$, $\delta = 0.98$, $a = 5$, $\epsilon = 0.05$, $S_{1} = 8$ $S_{2} = 4$, $1/Fr = 10^{-5}$, $1/Re = 10^{-4}$, $\theta = 10^{-4}$. As discussed in the main text, the discontinuity in the initial condition has clearly produced a more unstable flow, leading to the much earlier concentration of vorticity and development of the characteristic K-H structure.}
	\label{fig:2:vorticity_comparison}       
\end{figure*}

Figure \ref{fig:2:vorticity_comparison}  presents a comparison of the results of the discontinuous and smoothed initial velocity profiles, as vorticity contours taken at three select times $t = 1$, $t = 10$ and $t = 100$. The vorticity contours further demonstrate the nature of the instability, as it is clear from Figure \ref{fig:2:vorticity_comparison} that the discontinuous initial conditions derived from linear theory induce both an apparent physical instability and a clear numerical artefact, Gibbs' phenomenon, in the reconstructed vorticity. The former case, without recourse to further simulation or physical experiment, should be considered an effect of the latter. 

The case of hyperbolic tangent profile initial condition Equation (\ref{eq:continuous_velocity_ic}) in Figure \ref{fig:2:vorticity_comparison} (d) - (f) shows a refined confinement and constancy of vorticity within the inter-facial zone, while the discontinuous cases in Figure \ref{fig:2:vorticity_comparison} (a) - (c) show oscillations produced by Gibbs' phenomenon due to the discontinuity, of which the axial oscillations are of greater magnitude than those in the radial direction. The manifest confinement of vorticity to the inter-facial zone and the apparent constancy of the vorticity therein, suggests that the smoothed reconstruction of the vorticity has regularised the vortex sheet of the ideal theory such that the flow forms a shear layer. It is clear that the smoothed version of the initial condition on the axial velocity produces more stable results which take longer to develop the characteristic structure, as seen particularly in the last comparison in Figure \ref{fig:2:vorticity_comparison}, at simulation time $t = 100$. The density perturbation steepness parameter was set constant to $S_{2} = 4$ in order to model a smooth transition between the fluid regions. 

\subsection[]{Further Results}
\label{subsec:results_2}

\begin{figure*}
	\includegraphics[width=\columnwidth, scale=1]{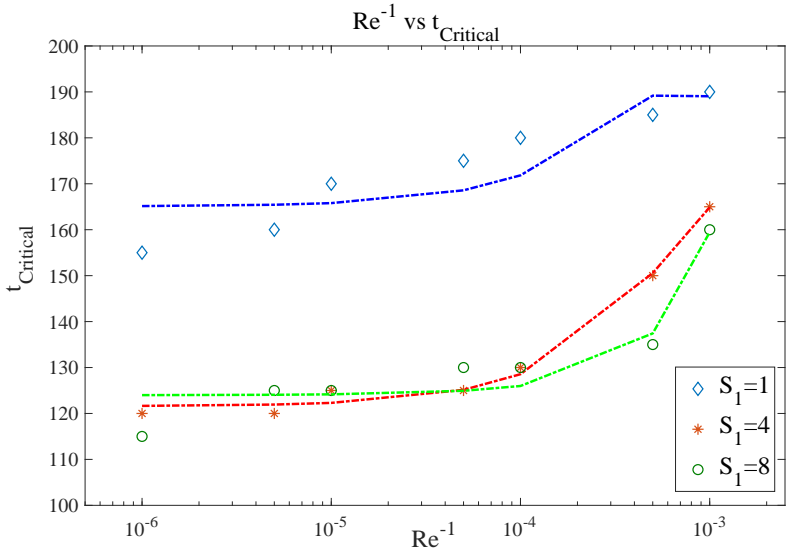}
	\caption{This log-linear plot presents a stability analysis of the numerical simulations. The  inverse Reynolds number of the flow is the $x$-axis of the graph. The $y$-axis is $t_{Critical}$ where this has been defined as the approximate on-set of non-linearity of the instability \ie when the interface of initially sinusoidal form begins to show asymmetric features. The other parameter is the steepness of the shear profile initial condition $S_{1} = 1,4,8$. The remaining parameters of the simulation were fixed at $N = 21$ (sufficient for these early times), $n = 155$, $\gamma = 0.95$, $\delta = 0.98$, $a = 5$, $\epsilon = 0.1$, $S_{2} = 4$, $1/Fr = 0$, $\theta = 5\times10^{-4}$. Imposed upon the points determined by visual inspection are least-squares lines-of-best fit.} 
	\label{fig:4:stability}       
\end{figure*}

For these results, the steepness parameter of the shear layer was generally set to $S_{1} = 8$ in order to model the ``top-hat" profile in which axi-symmetric modes are most unstable, \cite{batchelorgill1962}, or $S_{1} = 1$, as an approximation of the ``far-down-stream" profile. The research of \cite{batchelorgill1962} suggest that an inviscid jet with the ``top-hat" velocity profile is unstable to all modes of disturbance, while the ``far-downstream" profile is unstable exclusively to the first helical mode, which cannot be captured in this axi-symmetric model. The effect of this parameter is explored in a preliminary stability analysis of the numerical scheme presented in Figure \ref{fig:4:stability}. 

The inverse Reynolds number of the flow acts as abscissa of the graph, with corresponding ordinate of $t_{\text{Critical}}$ which is approximately defined as the (simulation) time at which the interface begins to over-turn \ie when non-linear instability becomes dominant. The points were determined by visual inspection of the simulation density and vorticity contours, to within a graphical accuracy of approximately 5 time units. Results are shown for the three different steepness parameters $S_{1} = 1$,$4$ and $8$. Polynomials of second-order as determined by an internal MATLAB least-squares method are superimposed to emphasise the general trend which suggests that the instability takes longer to develop in flows of higher inverse Reynolds number.

The magnitude of the inverse Reynolds number can be considered as quantifying the ratio of viscous to non-viscous forces on a fluid volume element, \cite{batchelor1970}. Thus the presented results suggest the fluid viscosity stabilises the flow, confirming the expected behaviour. The range $10^{-6}$ to $10^{-3}$ of the $x$-axis represents a portion of those values of the inverse Reynolds number explored in this research. 

Figure \ref{fig:4:stability} suggests that the effect of the steepness parameter of the initial axial velocity profile is to decrease the time of on-set of the instability of this axi-symmetric flow as the steepness increases. This corresponds well with intuition as well as with the results of the work on modelling jet spatial development, \eg \cite{batchelorgill1962}. As was suspected, the results suggest that this effect of the steepness parameter is negligible once the profile is ``steep enough". The Froude number and other parameters of the system are of secondary interest to the present research, and so have only been slightly explored and the results are not presented.

\begin{figure*}
	\includegraphics[width=\columnwidth, scale=1]{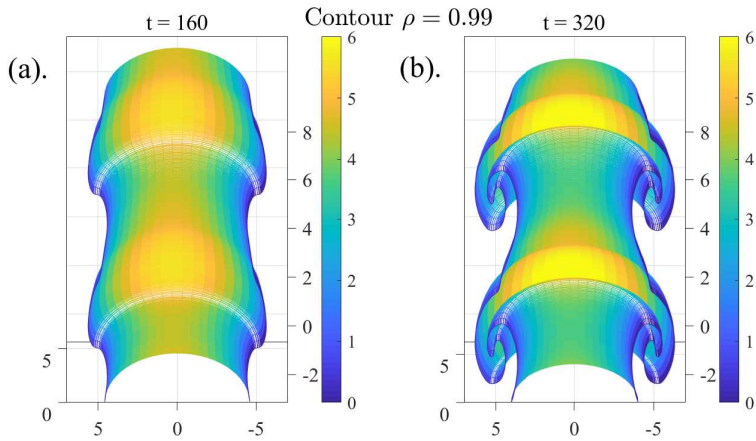}
	\caption{Cross-sections of  the $\rho = 0.99$ contour of a simulated fluid jet demonstrating the axial symmetry of the model, taken at the late times of $t = 160$ and $t = 320$ simulation units. The parameters used for these simulations are: $N = 41$, $n = 255$, $\gamma = 0.96$, $\delta = 0.98$, $a = 5$, $\epsilon = 0.1$, $S_{1} = 10$, $S_{2} = 10$, $1/Fr = 0$, $1/Re = 2\times10^{-4}$, $\theta = 2\times10^{-4}$. Each sub-figure is given a colour scale for the interface deflection to aid visualisation.}
	\label{fig:3:cut_away_jet}       
\end{figure*}

The initial stability of these results allowed the simulations to persist further into the non-linear stage of development. To demonstrate this, as well as the innate 3-dimensionality of these simulations, Figure \ref{fig:3:cut_away_jet} shows the $\rho = 0.99$ contour, half the initial density difference, as an approximation of a true interface, at the late times $t = 160$ and $t = 320$ simulation units, revolved around the  of symmetry half way to produce the jet as shown. As discussed above, the development of the instability causes the concentration of vorticity and over-turning of the interface. At the later time $t = 320$ shown in Figure \ref{fig:3:cut_away_jet}, the interface has almost spiralled around completely, forming the distinct  ``cat's-eye" structures of well-developed K-H instability, but here wrapped around the central jet as shown. There is a clear similarity to the observed phenomenon, \eg as exhibited in \cite{becker1968}. This later image also shows slight kinks on the closed side of each spiral, perhaps suggesting eventual vortex-shedding.

%

%

\section[]{Conclusion}
\label{sec:conclusion}

This work has explored the K-H instability as it occurs on the interfaces of round jets. As an initial analysis of the problem, the physics was idealised and studied in terms of equations linearised around the undisturbed jet. The results of this analysis agreed with the known results in the literature. To go beyond this precursory appraisal, a second model based upon the Boussinesq approximation was developed. On adopting the Boussinesq approximation the interface, a discontinuous density profile was smoothed into an inter-facial zone.  Furthermore, the curvature singularity discussed in the literature was avoided by this smoothing procedure.  Solutions to the Boussinesq model were sought using a computational spectral method. Simulations produced by this method were compared to the linearised prediction in the appropriate limit of the parameter space and found to be in good agreement at least for early times, as expected. Representative results showing the effects of finite-amplitude disturbances, beyond the early times predicted by the linear theory, were given. 

The effect of the parameters relevant to the initial velocity and vorticity of the fluid were explored. In particular, it has been found that the time to overturning of the interface is reduced when the initial shear magnitude and steepness in the profile is increased. Additionally, this time is also decreased when the density difference across the inter-facial region is increased.

It has been demonstrated that the fluid cylinder may evolve large-scale overturning portions, similar to the ``cat's-eye" spirals in planar K-H flow, except that here, they are arranged circularly around the central axis of the flow.

\begin{acknowledgements}
The authors gratefully acknowledge the scholarly assistance of Prof. A. Bassom and Dr S. Walters, as well as the financial assistance of the G.F.R. Ellis Memorial and Tasmania Graduate Research Scholarships. The critical comments from three Anonymous Reviewers improved the presentation and results of this paper significantly.

\end{acknowledgements}

%
\section*{Conflict of interest}

The authors declare that they have no conflict of interest.




\end{document}